\documentclass[preprint,12pt]{elsarticle}

\usepackage{graphicx}
\usepackage{amssymb}
\journal{Nuclear Physics B}

\def\barray{\begin{eqnarray}}
\def\earray{\end{eqnarray}}
\def\beq{\begin{equation}}
\def\eeq{\end{equation}}
\def\Rmath{{\bf R}}

\begin{document} 

\begin{frontmatter}

\title{An $xp$ model on AdS$_2$ spacetime}

\author[a]{Javier Molina-Vilaplana}
\author[b]{Germ\'an Sierra}
\address[a]{Universidad Polit\'ecnica de Cartagena, Cartagena, Spain}
\address[b]{Instituto de F\'{\i}sica Te\'orica,
UAM-CSIC, Madrid, Spain}

\begin{abstract}
In this paper we formulate the  $xp$ model on the AdS$_2$ spacetime. 
We find that the spectrum of the Hamiltonian has positive and negative
eigenvalues, whose absolute values are given by 
a  harmonic oscillator spectrum, which in turn   coincides with that of a massive
Dirac fermion in AdS$_2$.  We extend this  result to generic $xp$
models which are shown to be equivalent to a massive Dirac 
fermion on spacetimes whose metric depend of the $xp$ Hamiltonian. 
Finally, we  construct the generators of the isometry group $SO(2,1)$
of the AdS$_2$ spacetime,  and discuss the relation with conformal quantum mechanics. 

\end{abstract}


\end{frontmatter}

\section{Introduction}
\label{intro}
In 1999 Berry, Keating and Connes suggested  that a spectral realization of the Riemann zeros
could be achieved by quantizing the  classical Hamiltonian $H= xp$, where $x$
and $p$ are the position and momenta of a particle moving in one dimension \cite{BK99,C99}. 
Several recent  works have been devoted to clarify
the possible relation of the  $xp$ model with the Riemann zeros  \cite{ST08}-\cite{G12}.
In references \cite{SL11,BK11,S12}  it was advocated that one needs to modify the $xp$ Hamiltonian in order
to have a discrete spectrum, since the standard quantization of $xp$
yields a continuum \cite{S07a,TM07}.  The new Hamiltonians  take the form
$H = w(x) (p + \ell_p^2/p)$, where $w(x) = x \; ( x \geq \ell_x)$ and $w(x) = x + \ell_x^2/x \; ( x \geq 0)$,
with  $\ell_x \ell_p = 2 \pi \hbar$ \cite{SL11,BK11,S12}. 

 The latter  results led to the study of the  general Hamiltonian $H = w(x) (p + \ell_p^2/p)$
 for arbitrary positive functions $w(x)$ \cite{S12}. It turns  out that these models 
 describe the motion of a relativistic  particle moving in a 1+1 spacetime whose
 metric is determined by $w(x)$. The Riemann scalar curvature
 vanishes identically  for the linear  potential $w(x)=x$, 
and asymptotically for  $w(x) = x + \ell_x^2/x$.  
On the other hand, the potential 
$w(x) = w_0  \cosh(x/R)$ yields
a spacetime with constant negative curvature ${\cal R} = - 2/R^2$, which corresponds to 
an anti-de-Sitter spacetime  (AdS$_2$), with  radius $R$ \cite{S12}. 
Here though,   the semiclassical spectrum
has positive and negative branches each one given 
by a harmonic oscillator spectrum, so it is not connected with the spectral interpretation of the Riemann zeros \cite{S12}. 

 In this paper we shall focus on the $xp$ model whose underlying space-time is AdS$_2$. 
 We shall first show  that the previous semiclassical spectrum is exact in the quantum theory. 
 In the course of our investigation we found out that this spectrum coincides with the
 one of a  massive Dirac field  in AdS$_2$.  This connection is not special to
 AdS$_2$ but has a wider range of application. The general $xp$ model formulated in 
 reference \cite{S12} is  actually equivalent to a massive Dirac fermion on a space-time
 whose metric can be constructed out from the function $w(x)$ defining the $xp$ models. 
In this paper we shall give a proof of this unexpected connection between two a priori different theories. 
The AdS$_2$ metric has an isometry group given by   $SO(2,1)$, which led us to  construct
the generators of this group in the $xp$ model.   $SO(2,1)$ is also the conformal
 group in $0+1$ dimensions and is therefore the symmetry group of conformal quantum mechanics \cite{J72}-\cite{BE03}. 
This suggests a connection between field theories in  AdS$_2$ and a conformal 
quantum mechanics living on the one-dimensional boundary of AdS$_2$. 
However, to find  explicit realizations of this correspondence has shown to be rather elusive, motivating several  works to clarify the AdS$_2$/CFT$_1$ correspondence \cite{S98}-\cite{PL09}. At the end of the paper we shall present some remarks
concerning the conformal quantum mechanics formulated by Alfaro-Fubini-Furlan \cite{AFF} and related
studies by Jackiw and collaborators \cite{Ch11,JP}.  We have also included three appendices  with 
complementary material.

\vspace{3 cm} 

\section{The $xp$-AdS$_2$  model}

The general  $xp$ Hamiltonians  are defined by \cite{S12} 

\beq
H =  p \, U(x)  + \frac{\ell_p^2  \, V(x)}{p} , \qquad x \in \Rmath, 
\label{1}
\eeq
where $x$ and $p$ are the position and momentum of a particle moving in the real line $\Rmath$, 
$\ell_p$ is a parameter with the dimension
of momenta, and  $U(x)$ and $V(x)$ are positive functions. 
In references \cite{SL11,S12}, the  Hamiltonian
(\ref{1}) was  defined on a half-line, but this study we shall consider the whole
line  in order to discuss the AdS$_2$ model.

The Hamiltonian (\ref{1}) breaks time
reversal symmetry, under which $H \rightarrow - H$. 
Using the Hamiltonian equations of motion, the momentum $p$
can be expressed in terms of the velocity $\dot{x}$ as

\beq
p = \ell_p \, \eta \sqrt{  \frac{ V(x)}{ U(x) - \dot{x}}}
\label{p}
\eeq
where $\eta = \pm 1$ is the sign of the momentum and the energy, which are conserved quantities. 
Substituting (\ref{p})  back into the Lagrangian $L = p \, \dot{x} - H$ one obtains the action

\beq
S =  \int dt \, L = - \ell_p \, \eta  \int  \sqrt{ - ds^2}, 
\label{2}
\eeq
which, for either sign of $\eta$,  is the action  of a relativistic massive particle
moving in a spacetime with  metric 

\beq
 ds^2 = 4  V(x) ( - U(x)   dt^2  +   \, dt \, d x). 
\label{3}
\eeq
The momenta $\ell_p$ plays the role of $m c$, where $m$ is a mass and $c$ the speed of light. 
Note the special form of the metric where the component $g_{xx}$ vanishes. 
From eq.(\ref{1})
it is clear that the existence of positive and negative momentum implies that the energy also has
two possible signs, so that classically the spectrum consists  of two branches of energies.
Consequently, the semiclassical and quantum eigenergies will be positive and negative as well.

The Riemann scalar curvature is 

\beq
{\cal R}(x) = - \frac{ 1}{ V(x) }  \partial_x \left[ \frac{ \partial_x ( U(x) V(x)) }{ V(x) } \right], 
\label{4}
\eeq
such that  choosing $U$ and $V$ as  

\beq
U(x) = V(x) \equiv  w(x) = w_0 \, \cosh \frac{x}{R} \rightarrow {\cal R}(x) = - \frac{2}{ R^2}, \qquad x \in (- \infty, \infty). 
\label{5}
\eeq
gives a constant negative scalar curvature, that corresponds to a AdS$_2$ spacetime
of radius $R$.   This fact implies that  the classical trajectories generated by the classical Hamiltonian
(\ref{1}), can be seen as  the geodesics  of  the AdS$_2$ spacetime (see \ref{sec:tray}).  
The choice (\ref{5}) gives

\beq
\frac{1}{4} ds^2 = -  w_0^2  \cosh^2\left(  \frac{x}{R} \right)   dt^2 +   w_0 \cosh  \left( \frac{x}{R} \right)   dt  \, d x, 
\label{6}
\eeq
and making the change of variables

  \beq
 \sinh \frac{x}{R} = \tan \theta, \quad t = \frac{R}{ 2 w_0} ( \tau + \theta),  \quad \theta \in \left( - \frac{\pi}{2}, \frac{\pi}{2} \right), \quad \tau \in (- \infty, \infty) 
 \label{7}
 \eeq
brings  (\ref{6})  into an standard form of  the AdS$_2$  metric   (see \ref{sec:AdS}) 

 \barray 
ds^2  & = &    \frac{R^2}{ \cos^2 \theta} ( - d \tau^2 + d \theta^2). 
 \label{8}
\earray 


\section{Semiclassical spectrum}

The general covariance of the action (\ref{2}) allows one to choose a {\em symmetric} gauge where 
 $U(x)= V(x) \equiv w(x)$ \cite{S12}.  Let us further assume that $w(x)$ is an even function
 as in eq.(\ref{5}). In these cases 
the number of semiclassical energy levels, $n(E)$, between $0$ and $E>0$, is given by \cite{S12}

\beq
n(E) + \frac{1}{2}  = \frac{1 }{ 2 \pi \hbar} \int_{- x_M}^{x_M} \frac{dx}{ w(x)} \sqrt{E^2 - 4 \ell_p^2 w^2(x)} 
\label{12}
\eeq
where   $x_M$ is the turning point of the classical trajectories, i.e. $E = 2 \ell_p  w(x_M)$. The constant $\frac{1}{2}$
 has been included to  account for the  Maslow phase associated to the classical trajectories  (see \ref{sec:tray}). 
 Using the variable $\theta$ defined in eq.(\ref{7}) one finds 

\beq
n + \frac{1}{2} = \frac{2 R \ell_p}{ \pi \hbar} \int_{0}^{\theta_{M}} d \theta  \sqrt{\varepsilon^2 - \frac{1}{\cos^2 \theta} } 
= \frac{R \ell_p}{\hbar} ( \varepsilon-1), 
\label{13}
\eeq
where

\beq
\varepsilon = \frac{E}{2 w_0 \ell_p}  = \frac{1}{ \cos  \theta_M}. 
\label{14}
\eeq
Hence the semiclassical spectrum is given by

\beq
E_n = \frac{ 2 \hbar w_0}{ R} ( n + \frac{R \ell_p}{\hbar} + \frac{1}{2}), \qquad n=0,1, \dots
\label{15}
\eeq
which coincides with the harmonic oscillator spectrum with a zero point energy 
that  depends on the   dimensionless constant

\beq
\kappa =  \frac{R \ell_p}{\hbar},
\label{16}
\eeq
that  may  take any positive value. 
Note that the spectrum also contains  negative
eigenenergies whose absolute value is given by eq.(\ref{15}). 

\section{Quantum spectrum}

The classical Hamiltonian (\ref{1})  can be quantized in terms of the  following normal ordered operator \cite{S12} 

\beq
\hat{H} = \sqrt{U(x)}  \,    \hat{p} \,   \sqrt{U(x)}   +   \ell_p^2  \sqrt{V(x) }   \hat{p}^{-1}   \sqrt{V(x)}.
\label{17}
\eeq
where $\hat{p} = - i \hbar d/dx$ and $\hat{p}^{-1}$ is the pseudo-differential operator that
acts of wave functions as

\beq
(  \hat{p}^{-1}  \psi)(x) = - \frac{i}{\hbar}  \int_x^\infty dy \, \psi(y).
\label{17b}
\eeq
Note that $\hat{p} \, \hat{p}^{-1} = \hat{p}^{-1} \,  \hat{p} = {\bf 1}$ on wave functions that  vanish   at $x = \infty$.  
The normal ordering chosen in (\ref{17}) is not unique,  but it leads to an  hermitean operator under the conditions 
to be discussed below. In the case where $U(x) = x$ and $V(x) = 0$, equation (\ref{17}) yields
$\hat{H} = x^{1/2} \,  \hat{p}  \, x^{1/2} = \frac{1}{2} ( x \, \hat{p} + \hat{p} x)$, which coincides with the well known
Berry-Keating-Connes Hamiltonian \cite{BK99,C99}. 
Using  the previous definitions the action of (\ref{17}) is  given by

\beq
( \hat{H} \psi)(x) = - i \hbar \sqrt{U(x)} )\frac{d}{dx} \{ \sqrt{U(x)} \psi(x) \} 
-   \frac{i \ell_p^2}{ \hbar}  \int_{x}^\infty dy \, \sqrt{V(x) V(y)} \psi(y). 
\label{19}
\eeq
$\hat{H}$  is a symmetric operator, i.e.

\beq
\langle \psi_1 | \hat{H} \psi_2 \rangle = \langle \hat{H} \psi_1 |\psi_2 \rangle = 0,
\label{20}
\eeq
acting on wave functions that   satisfy  the   conditions

\beq
\lim_{x \rightarrow \pm \infty} \sqrt{U(x)}  \psi(x) =0, \qquad \int_{- \infty}^\infty dx \, \sqrt{V(x)} \, \psi(x) = 0. 
\label{21}
\eeq
Choosing  the symmetric gauge and defining  the function

\beq
u(x) = \sqrt{ w(x)}, 
\label{18}
\eeq
one can write 
the Schroedinger equation  as

\beq
 - i \hbar u(x)\frac{d}{dx} \{ u(x) \psi(x) \} 
-  \frac{i \ell_p^2}{ \hbar}  \int^{ \infty}_x dy \, u(x)  \, u(y) \psi(y)  = E \psi(x), 
\label{22}
\eeq
and similarly 
\beq
 \hbar^2 \frac{d}{dx} \{ u(x) \psi(x) \} +  \ell_p^2 
 \int^{\infty}_x dy   \, u(y) \psi(y)  - i E \hbar  \frac{ \psi(x)}{ u(x)}   = 0. 
\label{23}
\eeq
In terms of the function 

\beq
\phi(x) = u(x) \psi(x), 
\label{24}
\eeq
eq.(\ref{23})  reads

\beq
 \hbar^2 \frac{d  \phi(x)}{dx}  +  \ell_p^2 
 \int^{\infty}_x d y   \,   \phi(y)  - i E \hbar  \frac{ \phi(x)}{ w(x)}   = 0, 
\label{25}
\eeq
and  the conditions  (\ref{21}), 

\beq
\lim_{x \rightarrow \pm \infty} \phi(x) =0, \qquad \int_{- \infty}^\infty dx \, \, \phi(x) = 0. 
\label{26}
\eeq
To solve eq. (\ref{25}) one takes a  derivative   obtaining  

\beq
 \hbar^2 \frac{d^2 \phi(x)}{dx^2}  
 - i E \hbar  \frac{d}{dx} \left(   \frac{ \phi(x)}{ w(x)}  \right) - \ell_p^2 \phi(x)   = 0. 
\label{27}
\eeq
Equations (\ref{27}) and (\ref{26}) imply   (\ref{25})  for functions $\phi(x)$ which decay sufficiently fast at $\pm \infty$.
For the function 

\beq
w(x) = w_0 \cosh \frac{x}{R}, 
\label{28}
\eeq
one gets

\beq
\partial_x^2 \phi(x) - \frac{ i E }{ \hbar w_0 } \frac{1}{ \cosh(x/R)} \partial_x \phi(x) + \left( 
\frac{ i E }{ \hbar w_0   R} \frac{ \sinh(x/R)}{ \cosh^2(x/R)} - \frac{\ell_p^2}{ \hbar^2}  \right) \phi(x) = 0. 
\label{29}
\eeq
In the limit $|x| \rightarrow \infty$ this equation becomes 

\beq
\partial_x^2 \phi(x) -  \frac{\ell_p^2 }{ \hbar^2}   \phi(x) \sim  0, 
\label{30}
\eeq
so that the wave function decays  asymptotically as

\beq
\phi(x) \rightarrow e^{ - |x| \ell_p/\hbar}, \qquad |x| \rightarrow \infty
\label{31} 
\eeq
To solve eq.(\ref{29}) we first write it in terms of the angle variable $\theta$, 

\beq
\cos^2 \theta \;  \partial_\theta^2 \phi -  \cos \theta( \sin \theta + i  \alpha \cos \theta ) \partial_\theta \phi + 
( i \alpha \sin \theta \cos \theta  - \kappa^2) \phi = 0
\label{32}
\eeq
where

\beq
\alpha = \frac{E R}{ \hbar w_0}.
\label{33}
\eeq
Notice that the semiclassical result (\ref{15}) implies for $\alpha$

\beq 
\alpha^{\rm (sc)}_n = 2 n + 2 \kappa +1, \qquad n=0,1, \dots
\label{33b}
\eeq
together with the negative values $- \alpha^{\rm (sc)}_n$. 
Next we define the function $f(\theta)$ as

\beq
f(\theta) = 2 \cos \theta \;  \phi(\theta),  
\label{34}
\eeq
which satisfies

\beq
\cos^2 \theta \;  \partial_\theta^2 f  +   \cos \theta( \sin \theta -  i  \alpha \cos \theta ) \partial_\theta f + 
(1  - \kappa^2) f = 0.
\label{35}
\eeq
Defining the complex variable 

\beq
z = e^{ 2  i \theta},
\label{36}
\eeq
eq. (\ref{35}) becomes

\beq
z ( z+1)^2  \; \partial_z^2  \, f  +  \frac{1}{2} (z+1) [ (1- \alpha ) z + 3 - \alpha] \,  \partial_z \, f  + (\kappa^2-1) f  = 0,
\label{37}
\eeq
One can  verify  that if $f(z)$ is a solution of this equation  then $f(z^{-1})$ is a solution with $\alpha$ replaced
by $- \alpha$. In this manner one gets the negative energy solutions from the positive ones. 
The conditions  (\ref{26}) read

\barray
\lim_{ z \rightarrow e^{ \pm i \pi}} \frac{ f(z)}{ (1+z)} & = & 0, \label{38} \\
\int_{\cal  C}  dz \, \frac{f(z)}{ (1+z)^2}  & = &0. \label{39}
\earray 
where ${\cal C}$ is the unit circle $|z|=1$.  Choosing  $\kappa=1$, the solution of (\ref{37}) can be easily be found

\beq
f(z) = A z^{\frac{ \alpha -1}{2}} \left( \frac{z}{ \alpha +1} + \frac{1}{ \alpha-1} \right) + B, \quad \alpha \neq 1, -1. 
\label{40}
\eeq
where $A$ and $B$ are integration constants. 
The values $\alpha =1, -1$ are excluded because the associated   $f(z)$ functions  contain  $\log z$ such  that  the condition (\ref{38})
is  not satisfied. From (\ref{38})  and (\ref{40}) one finds

\beq
f( e^{ \pm i \pi}) = 0 \rightarrow B = \frac{2 A}{1- \alpha^2} e^{ \pm i \frac{ \pi}{2} ( \alpha-1)} \rightarrow 
e^{ i \pi \alpha} = -1
\label{41}
\eeq
which yields the quantization conditions 

\beq
\alpha_{n, \pm} =
\left\{ \begin{array}{lc}
( 2 n +3), \qquad n=0,1, \dots, & \alpha >  0 \\
 ( 2 n +3), \qquad n=-3,-4, \dots, & \alpha <  0 \\
 \end{array}
 \right. 
\label{42}
\eeq
that  coincide  with the semiclassical result (\ref{33b}) for $\kappa=1$. 
The   expression (\ref{40}) for the positive energy solutions is 

\barray
f_{\kappa= 1, n,+}(z) & = & (-1)^n \left[ (n+1) z^{n+2}  + (n+2)  z^{n+1} + (-1)^n \right] \label{43} \\
\nonumber
\earray 
and for the negative energy solutions 

\beq
f_{\kappa= 1, -n-3,-}(z)= (-1)^n  \frac{  n+1}{n}  
f_{\kappa= 1, n,+}(z^{-1}) , \qquad n=0, 1, \dots
\label{44}
\eeq
The factor multiplying $f_{\kappa= 1, n,+}(z^{-1})$ will be explained below. 
Finally, to verify the condition (\ref{39}),  we express (\ref{43}) as

\barray 
f_{\kappa= 1, n,+}(z)
& = &  (z+1)^2 \, \sum_{r=0}^n  (r+1) (-z)^r.  \label{46}
\earray  
The vanishing of the integral follows from the Cauchy theorem. 
The solutions $f_{\kappa= 1, n,-}(z)$ also satisfy eq.(\ref{39}), which  can be written as

\barray
\int_{\cal C}  dz \, \frac{f(z^{-1})}{ (1+z)^2}  & = &0. \label{47}
\earray 

Let us next  consider generic values of $\kappa$. Eq.(\ref{37}) has two linear independent solutions given by

\barray
f_{\kappa, n, +}(z) &=& (z+1)^{\kappa+1} \, F(\kappa +1, -n, - \kappa -n+1, -z),  \label{48} \\
f_{\kappa, n, -}(z) &=& (z+1)^{\kappa+1} \, z^{ \kappa+n} \,  F(\kappa ,2 \kappa +  n +1,  \kappa +n+1, -z) \nonumber 
\earray 
where $F$ is the  hypergeometric function of type $F_{2,1}$,  and 
$\alpha$ has been parametrized as

\beq
\alpha = 2 n + 2 \kappa +1, 
\label{49}
\eeq
with  $n$ an arbitrary number.  One can transform  these eqs. into  \cite{AS} 

\barray
f_{\kappa, n, +}(z) &=& (z+1)^{1- \kappa} \, F(- 2 \kappa -n , - \kappa + 1, - \kappa -n+1, -z),  \label{50} \\
f_{\kappa, n, -}(z) &=& (z+1)^{1- \kappa} \, z^{ \kappa+n} \,  F(n+1 , - \kappa,  \kappa +n+1, -z).  \nonumber 
\earray 
The Gauss series defining these  hypergeometric functions are absolutely convergent
in the unit cycle $|z|=1$ \cite{AS}. The condition (\ref{38}) implies

\barray
F(- 2 \kappa -n , - \kappa + 1, - \kappa -n+1, 1) &=&  \frac{ \Gamma(2 \kappa) \Gamma(- \kappa - n +1)}{  \Gamma( \kappa+1) \Gamma(  - n)} = 0,  \label{51} \\
 F(n+1 , - \kappa,  \kappa +n+1, 1)  &=&  \frac{ \Gamma(2 \kappa) \Gamma( \kappa + n +1)}{  \Gamma( \kappa) \Gamma( 2 \kappa + n+ 1)}  = 0. \nonumber 
\earray 
For generic values of $\kappa$, the first of these eqs. implies that $n=0, 1, \dots$, and correspond  to the positive energy solutions,
while the second eq. implies $n = -( 2 \kappa + p +1) \; (p=0,1, \dots)$, and  correspond to the negative energy solutions
$\alpha = - ( 2 p + 2 \kappa+1)$.   The relation between both solutions  is given by 

\beq
f_{\kappa, -n - 2 \kappa -1, -}(z) = (-1)^n \frac{ \kappa+n}{ \kappa}
f_{\kappa, n, +}(z^{-1}) 
, \; \; n=0,1, \dots.
\label{51b}
\eeq
and can be proved using hypergeometric identities \cite{AS}.  The case $\kappa=1$ reproduces eq.(\ref{44}). 
As follows from (\ref{48}),  the functions $f_{\kappa, n, +}(z)$ 
are the product of a polynomial of degree $n$ times the factor $(z+1)^{ \kappa+1}$. Some examples are

\barray
f_{\kappa, 0, +}(z) & = &  (z+1)^{\kappa +1}  \label{52} \\
f_{\kappa, 1, +}(z) & = &  (z+1)^{\kappa +1}   (1 -  ( 1+ \kappa^{-1} ) z)  
\nonumber \\
f_{\kappa, 2, +}(z) & = &  (z+1)^{\kappa +1}   (1 -  2 z + ( 1+ 2 \kappa^{-1} ) z^2)  
\nonumber \\
\dots & & \nonumber 
\earray 
which for  $\kappa=1$, reproduces eqs.(\ref{46}).  Finally, the condition (\ref{39}) is also satisfied  by  the Cauchy theorem. 

The wave functions $\psi(x)$ can be written as (recall eqs.(\ref{24}) and  (\ref{34}))

\beq
\psi = (\cos \theta)^{1/2} \, \phi(\theta) = \frac{ f(\theta)}{ 2  (\cos \theta)^{1/2}}, 
\label{53}
\eeq
and the scalar product as 

\barray 
\langle \psi_1 | \psi_2 \rangle & = &  \int_{- \infty}^\infty dx \, \psi_1^*(x) \psi_2(x) = R 
\int_{- \frac{\pi}{2}}^{\frac{\pi}{2}} d \theta \, \phi_1^*(\theta) \phi_2(\theta)  \label{54} \\
& = &  
\frac{R}{4}  
\int_{- \frac{\pi}{2}}^{\frac{\pi}{2}}  \frac{d \theta}{ (\cos \theta)^2}  \, f_1^*(\theta) f_2(\theta)  =
\frac{R}{2 i}  
\int_{\cal C}  \frac{d z}{ (z+1)^2}  \, f_1^*(z) f_2(z).
\nonumber
\earray 
Using this formula one can compute the norm of the wave functions. For example for $\kappa=1$ 
one gets

\beq
\langle \psi_{\kappa=1, n, +}  | \psi_{\kappa=1, n, +}  \rangle = \pi R (n+1)(n+2), \quad n=0,1, \dots. 
\label{55}
\eeq

\section{The general $xp$ model as a massive Dirac fermion in space-time}

The spectrum obtained above turns out to coincide with the spectrum 
of a massive  Dirac  fermion in  AdS$_2$, which we reproduce for completeness  in Appendix C (see 
\cite{Dirac,Villa}).  
This result may look surprising given 
that  the Dirac fermion in 1+1 dimensions is a two component spinor,  $\psi_\pm$, while our $xp$ model
involves  a single  wave function $\psi$. The relation between the $xp$ model and the massive Dirac fermion
in AdS$_2$ is indeed more general and not specific to the AdS$_2$ space-time. In this section
we shall derive this relation in detail which comes from   the special 
form  of the  spacetime metric (\ref{3}) underlying the $xp$ model.

Let us start from the massive Dirac equation 
in 1+1 spacetime dimensions
\beq
( e^\mu_a \gamma^a D_\mu +   \frac{i m c}{\hbar} ) \psi = 0,
\label{e16}
\eeq
where $\psi^t= (\psi_-, \psi_+)$ is a two component spinor, 
 $e^\mu_a$ is  the inverse of the vielbein $e^a_\mu$, 
 $\gamma^a$ are  given  in terms
 of the Pauli matrices as  $\gamma^0 = \sigma^x, \gamma^1= - i \sigma^y$, 
 $D_\mu$ is the covariant derivative, 
 \beq
D_\mu \psi = (\partial_\mu -  \frac{1}{4} \omega_\mu^{ab} \gamma_{ab} ) \psi, 
\label{e15}
\eeq 
 where $\omega_\mu^{ab}$ is the spin connection
 
  \beq
 \omega_\mu^{ab} = e^a_\nu \partial e^{b \nu} + e^a_\nu e^{b \lambda } \Gamma_{\lambda \mu}^\nu
 \label{d4}
 \eeq
 with $\Gamma_{\lambda \mu}^\nu$ the Christoffel symbols,  and 
  $\gamma_{ab} = \frac{1}{2} [ \gamma_a, \gamma_b]$.   
The   vielbein $e_\mu^a$  corresponding  to the metric (\ref{3}), that is
$g_{\mu \nu} = \eta_{ab} e^a_\mu e^b_\nu \;$ with ${\rm diag} \;  \eta_{ab} = (-1,1)$, 
can be chosen as 

\barray
e_0^0 & = &  V + U, \quad e_0^1 = V- U, \quad  e_1^1 = - e_1^0 =  1,  \label{e18} 
\earray
which yields  the  spin connection

\barray
\omega_0^{01} & = & \frac{ 1}{V} \partial_x ( U V), \quad \omega_1^{01}  =  - \frac{ 1}{V} \partial_x V.
\earray 
Plugging these expressions into the Dirac equation  (\ref{e16}) yields
 
\barray 
 \frac{1}{\sqrt{V}}  \,  \partial_x \left(  \sqrt{V}  \psi_+ \right) - \frac{i  m c}{ \hbar } \psi_-   & = & 0,  \label{e19} \\
\partial_t \psi_-  + \sqrt{U}  \partial_x  \left(  \sqrt{U}  \psi_-  \right)  + \frac{ i m c}{\hbar} V  \psi_+ & = & 0,  \label{e20}
\earray 
which do  not contain the time derivative of $\psi_+$. This feature implies that  the 
field $\psi_+$ can be expressed in terms of $\psi_-$ as 

\beq
\psi_+(x,t) = -  \frac{i m c/\hbar}{ \sqrt{ V(x)}} \int_{x}^\infty dy \, \sqrt{ V(y)}  \, \psi_-(y,t), \label{e21}
\eeq
under the asumption    $\lim_{x \rightarrow - \infty} \psi_+(t, x) = 0$. 
Replacing (\ref{e21}) into (\ref{e20}) yields the equation of motion of $\psi_-$

\beq 
i \partial_t \psi_-  = - i  \sqrt{U}  \partial_x  \left(  \sqrt{U}  \psi_-  \right)  - i 
\left( \frac{m c}{ \hbar} \right)^2   \sqrt{ V(x)}  \int_{x}^\infty dy \, \sqrt{ V(y)}  \, \psi_-(y,t) 
\label{e23} 
\nonumber 
\eeq
which  is nothing but  a Schroedinger equation  for $\psi_-$ with  the  Hamiltonian  (\ref{19}) under  the identifications
$\psi_- = \psi$ and 

\beq
\ell_p = m c.
\label{e26}
\eeq
Thus the origin of the  non local  term $\hat{p}^{-1}$ in (\ref{19}),
comes from the non dynamical  nature   of the field 
$\psi_+$, which in turn is due to the special form of the metric (\ref{3}). 
It would be interesting to investigate which are the spacetimes admitting   a
$xp$ version along the lines shown above. A condition would  be 
space-times with  time like killing vectors, that  will give rise
to a Hamiltonian evolution.

\section{$SO(2,1)$ symmetry}

Let us now return to the AdS$_2$ model. Here we notice that the 
 AdS$_2$  metric  has the isometry group $SO(2,1)$. We then  expect  that this
symmetry group can be realized in the  $xp$-AdS$_2$ model. In this section
we shall  construct the generators of this group, first in the classical
theory and then in  quantum mechanics. The group $SO(2,1)$
is locally  isomorphic to $SU(1,1)$ whose generators, $L_n \; (n=0, \pm1)$,  satisfy
the Poisson brackets

\beq
\{  L_0, L_{\pm 1} \}  = \pm   i \, L_{\pm 1},   \qquad \{  L_1, L_{-1} \}  =- 2   i  \, L_{0}.
\label{57}
\eeq
Choosing $L_0$  to be proportional to the classical Hamiltonian (\ref{1}), i.e. 

\beq
L_0 = \frac{R}{ 2 w_0} H, 
\label{59}
\eeq
one finds
 
 \barray
L_0 &=& \frac{R}{2} \cosh \left(  \frac{x}{R} \right)  \left( p + \frac{\ell_p^2}{p} \right),  \label{58} \\
L_{\pm 1}  &=& \frac{R}{2} \cosh \left(  \frac{x}{R} \right)  \left( p  \, e^{ \mp 2 i  \tan^{-1} (\sinh(x/R))}   -  \frac{\ell_p^2}{p} \right). 
\nonumber 
\earray 
Upon quantization the algebra (\ref{57}) becomes

\beq
[   \hat{L}_0, \hat{L}_{\pm 1} ]   = \mp \hbar  \, \hat{L}_{\pm 1},   \qquad [  \hat{L}_1, \hat{L}_{-1} ]   = 2   \hbar   \, \hat{L}_{0}, 
\label{60}
\eeq
which  holds  by a normal ordered version  of (\ref{58})

 \barray
\hat{L}_0 &=& \frac{R}{2} u(x)  \left(  \hat{p}  + \frac{\ell_p^2}{ \hat{p}} \right) u(x), \quad u(x) =   ( \cosh(x/R))^{1/2}, 
 \label{61} \\
\hat{L}_{\pm 1}  &=& \frac{R}{2} u(x)  \left(   e^{ \mp  i  \tan^{-1} (\sinh(x/R))}  \hat{p}  \;   e^{ \mp  i  \tan^{-1} (\sinh(x/R))}   -  \frac{\ell_p^2}{ \hat{p}} \right) u(x).
\nonumber 
\earray 
To verify eqs.(\ref{60}),   it is convenient   to perform  the similarity  transformation 

\beq
\hat{A'} = u \,  \hat{A} \;  u^{-1}, 
\label{62}
\eeq
which amounts  to act on the wave function $\phi$ related to $\psi$ by  eq.(\ref{24}), i.e.

\beq
\hat{A} \, \psi = u^{-1} \hat{A'} \, \phi. 
\label{63}
\eeq
The transformation of  $\hat{L}_n$  under (\ref{62}) is given in  the variable $\theta$ by

\barray 
(\hat{L'_0}  \phi) ( \theta)   & = &  - \frac{i \hbar}{2} \partial_\theta \phi(\theta)    - \frac{i \hbar \kappa^2}{2} \frac{1}{ \cos \theta}
\int_{\theta}^{ \pi/2} \frac{ d \theta'}{ \cos \theta'} \phi( \theta'),  
\label{64}  \\
(\hat{L}'_{\pm 1}   \phi) ( \theta)   & = &  - \frac{i \hbar}{2} e^{ \mp 2 i \theta} \, ( \partial_\theta \mp i)  \phi(\theta)    +  \frac{i \hbar \kappa^2}{2} \frac{1}{ \cos \theta}
\int_{\theta}^{ \pi/2} \frac{ d \theta'}{ \cos \theta'} \phi( \theta'). \nonumber   
\earray 
The eigenfunctions of the Hamiltonian $\hat{H}$ imply 

\beq
\hat{L'_0} \,  \phi_{\kappa, n, \pm } =  \pm \hbar ( n + \kappa + \frac{1}{2} )  \phi_{\kappa, n, \pm}, \qquad n=0,1, \dots. 
\label{65}
\eeq
Moreoever,  the positive and negative energy solutions satisfy

\beq
\hat{L}'_{\pm 1}  \,  \phi_{\kappa, 0, \pm } =  0. 
\label{65b}
\eeq
so that the  positive energy eigenfunctions can be obtained acting on $\phi_{\kappa, 0, +}$ with the raising operator 
$(\hat{L}'_{- 1})^n$. Similarly,  the negative energy  eigenfunctions  can be  constructed 
acting on  $\phi_{\kappa, 0, -}$  with  $(\hat{L}'_{1})^n$ . These two   infinite
dimensional representations of $SO(2,1)$ are related by complex conjugation. 

\section{The $xp$-AdS$_2$ model and  conformal quantum mechanics}

The group $SO(2,1)$ describes  the symmetry  of conformal quantum mechanics (CFT$_1$). 
The most studied   model  of a CFT$_1$ was introduced by Jackiw  in 1972  \cite{J72},  and its properties 
were analyzed in great detail  by de Alfaro, Fubini and Furlan (dAFF) \cite{AFF}.
The dAFF  model has been recently discussed in the framework of the  AdS$_2$/CFT$_1$ correspondence \cite{Ch11,JP,S98} and can be considered as the conformal limit of the matrix model describing flux backgrounds of 2D type 0A string theory \cite{strominger2}. It also has been shown that the spectrum of this limit is similar to that of a free fermion on AdS$_2$ \cite{Dirac}.

In this section we shall discuss  the relation of these works with the $xp$-AdS$_2$ model,  which in turn may shed
new light into the AdS$_2$/CFT$_1$ correspondence. 

The dAFF Hamiltonian describes  a particle on a 
 half-line subject to an inverse square interaction potential, i.e. 
 
 \beq
 H  = \frac{1}{2} ( p^2 + \frac{g}{x^2}), \quad x > 0, \qquad g >0, 
 \label{66}
 \eeq
 which together with the  operators $D$ and $K$, defined as
 
 \barray
 D & = & t H - \frac{1}{4} ( x p + p x),  \label{67} \\
 K & = & - t^2 H + 2 t D + \frac{1}{2} x^2,   \nonumber 
 \nonumber 
 \earray 
 generate  the   $SO(2,1)$ algebra \cite{AFF,Ch11} 
 
 \beq
 i[ D, H] = H, \quad  i[ D, K] = -K , \quad   i[ K, H] = 2D.
 \label{67b}
 \eeq
 The Hamiltonian $H$ is essentially self-adjoint for $g \geq 3/4$ with no discrete spectrum, 
 while for  $- \frac{1}{4} < g < \frac{3}{4}$  it  admits self-adjoint extension and has
 a bound state of negative energy (see \cite{BG,susy} and references therein). 
 Herethough  we shall consider not the spectrum of $H$ but that of an operator
 belonging to the  $SO(2,1)$ algebra  which in the Cartan basis  (\ref{60}) is given by 
 \cite{AFF,Ch11}

 \barray 
 L_0  &  = &  \frac{1}{2} \left( \frac{K}{a} + a H \right),   \label{68} \\
 L_{\pm 1}   &  = &  \frac{1}{2} \left( \frac{K}{a} -  a H \right)  \mp i D,  
\nonumber
\earray  
 where   $\hbar a$ has dimensions of (length)$^2$,
and such that $L_0$ has a discrete spectrum, i.e. 

\barray 
L_0 \, | n \rangle  =  \hbar  \, r_n \,  |n \rangle, \label{69} \\ 
r_n =   n +  r_0,    \quad n=0, 1, \dots,    \quad   r_0 >0,  \nonumber \\
 \langle n | n' \rangle   =    \delta_{n,n'}.  \nonumber 
\earray
The action of the operators $L_{\pm 1}$ acting in  this   basis is 

\beq
L_{\pm 1} \, |n \rangle =  \hbar \sqrt{ r_n ( r_n \mp 1) - r_0 ( r_0 -1)} \, |n \mp 1 \rangle. 
\label{70}
\eeq 
The parameter $r_0$ labels the representation of the algebra and is related to the Casimir ${\bf L}^2$ 
of $SO(2,1)$ as

\beq
{\bf L}^2  \, |n \rangle = \hbar^2  \,  r_0 ( r_0 -1) \, |n \rangle, \quad {\bf L}^2  = L_0^2  - \frac{1}{2} ( L_{1} L_{-1} +   L_{-1} L_1).
\label{70b}
\eeq
In the dAFF model,  $r_0$ is given by 

\beq 
r_0  = \frac{1}{2} \left( 1 + \sqrt{ g + \frac{1}{4}}  \right). 
\label{71}
\eeq
On the other hand,  the  value of $r_0$ in the $xp$-AdS$_2$ model 
is given by (see (\ref{65})) 

\beq
r_0 = \kappa + \frac{1}{2}.  
\label{72}
\eeq
Although the dAFF and the  $xp$-AdS$_2$ are two different models,  one can establish  a 
 correspondence between their parameters  based on dimensional arguments  and their  spectrum 

\beq
\frac{1}{2} \sqrt{g + \frac{1}{4}} \leftrightarrow \frac{  R \ell_p}{\hbar}, \qquad  \hbar \, a \leftrightarrow R^2, 
\label{73}
\eeq
which links $a$ to the radius of the AdS$_2$ space,  and $g$ to that radius measured
in units of the length $\hbar/ \ell_p$. However, unlike the dAFF model, where $g$ and $a$ 
does not make direct reference to the AdS$_2$ space-time, the parameters of the $xp$
model do: $R$  is  the radius of AdS$_2$  and 
$\ell_p/c$ is  the mass of the  fermion in the Dirac formulation
of the $xp$ model.  We hope that the results presented in this article may shed 
new  insights  on the AdS$_2$/CFT$_1$  correspondence. 

 \section*{Acknowledgements}
The authors acknowledge discussions with Esperanza L\'opez, J.L.F Barb\'on and Bernard Julia. This work has been financed by the 
 grants FIS2012-30625, Fundaci\'on S\'eneca Regi\'on de
Murcia 11920/PI/09,  FIS2012- 33642, QUITEMAD, and the Severo Ochoa Program.

\appendix
\section{Short review on  the AdS$_2$  spacetime}
\label{sec:AdS}

The AdS$_2$ spacetime is the locus  of  the one-sheeted  hyperboloid  \cite{ads} 

\beq
X_0^2 + X_2 ^2 - X_1^2 = R^2, 
\label{a1}
\eeq
where $R$  is a length denoted  the radius of AdS. The spacetime metric 
is inherited from the ambient  Minkowski  spacetime, 

\beq
(ds)^2 = - d X_0^2 - d X_2^2 + d X_1^2, 
\label{a2}
\eeq
which shows that  $SO(2,1)$ is  the isometry group, whose compact subgroup
$SO(2)$ can be identified with time.  The hyperboloid (\ref{a1}) can be described in  global coordinates as 

\barray
X_0  & = & R \, \cosh \rho  \; \cos \tau,  \label{a3}  \\
X_2  & = & R \, \cosh \rho  \; \sin \tau,   \nonumber \\
X_1  & = & R \, \sinh  \rho,   \; \nonumber 
\earray 
where  $\rho \in (- \infty, \infty)$ and $\tau \in [0, 2 \pi)$. The universal covering of AdS$_2$ 
spacetime  is obtained  letting  $\tau$ to take any real value.  The metric in these coordinates is

\beq
ds^2 = R^2 ( - \cosh^2 \rho \, d \tau^2 + d \rho^2 ). 
\label{a4}
\eeq
 The boundary at infinity, $\rho  =  \pm \infty$,   consists of  two disconnected worldlines 
 parameterized by the time coordinate   $\tau \in (- \infty, \infty)$.  The coordinate $\rho$ is identified
 with $x/R$ in the $xp$-AdS$_2$ model.   Throughout this paper 
 we have used  the variable  $\theta$ defined as 

\beq
\sinh \rho = \tan \theta, \qquad \theta \in (- \frac{\pi}{2}, \frac{ \pi}{2} ).
\label{a5} 
\eeq
 The boundaries of the spacetime are at  $\theta = \pm \pi/2$.  This coordinate 
 displays clearly  the causal structure of the metric (\ref{a4}), i.e. 

\beq
ds^2 = \frac{R^2}{ \cos^2 \theta} ( - d \tau^2 + d \theta^2). 
\label{a7}
\eeq

\section{Classical trajectories and geodesics}
\label{sec:tray}

The equations  of motion associated to the Hamiltonian (\ref{1}) are  \cite{S12}

\barray
\dot{x} &=& w(x) ( 1- \frac{\ell_p^2}{p^2}),  \label{b19} \\
\dot{p} &=& - w'(x) ( p + \frac{\ell_p^2}{p}).   \nonumber
\earray 
The energy $E$ is a conserved quantity, 

\beq
E = w(x) ( p + \frac{\ell_p^2}{p}), \label{b20}
\eeq
as well as the sign of the momenta $p$, which coincides with the sign of $E$ since
$w(x) >0,  \; \forall x$. We shall assume below that $E, p>0$. For each position $x$
there are two possible values of the momenta, 

\beq
p_\eta(x, E) = \frac{1}{2 w(x)} ( E + \eta  \sqrt{ E^2 - (2  \ell_p   w(x))^2}), \qquad \eta = \pm 1, 
\label{b21}
\eeq
related by 

\beq
p_-(x, E) = \frac{\ell_p^2 }{ p_+(x,E)}. 
\label{b22}
\eeq
The classical allowed region is given by

\beq
E \geq 2 \ell_p  w(x).  
\label{b23}
\eeq
Replacing  (\ref{b21}) into (\ref{b19}) yields 

\barray
\dot{x} &=& \frac{1}{2 \ell_p^2  w(x)}  \left( ( 2 \ell_p  w(x))^2 - E^2 + \eta E \sqrt{E^2 - ( 2 \ell_p  w(x))^2 }  \right), 
\label{b24} 
\earray 
so that the classical trajectories are given by

\beq
\int_{x_0}^x  dx \;  \frac{2 \ell_p^2  w(x)}{  ( 2 \ell_p  w(x))^2 - E^2 + \eta  E \sqrt{E^2 - ( 2 \ell_p w(x))^2 }} = t - t_0.  
\label{b25}
\eeq
To integrate this eq. we use the variable $\theta$ defined in (\ref{7}) or (\ref{a5})  and 

\beq
w(x) = w_0 \, \cosh \frac{x}{R} =  \frac{w_0}{ \cos \theta}. 
\label{b26}
\eeq
After some algebra (\ref{b25}) becomes (we choose $x_0=t_0$ so that $\theta_0=0$) 

\beq
\frac{R}{2 w_0} 
\int_0^{\theta}   d \theta  \;  \left[ 1 + \frac{ \eta  \, \varepsilon  \cos \theta}{ \sqrt{ \varepsilon^2 \cos^2 \theta -1}}  \right]   = t, 
\label{b27}
\eeq
where we have defined (recall (\ref{14})) 

\beq 
 \varepsilon  = \frac{E}{ 2 w_0 \ell_p}    \geq  1. 
\label{b28}
\eeq
Performing the integral one finds

\beq
\theta + \eta  \tan^{-1}  \left[  \frac{  \, \varepsilon  \sin  \theta}{ \sqrt{ \varepsilon^2 \cos^2 \theta -1}}  \right] = \frac{2 w_0 t}{ R} =
 \tau + \theta. 
\label{b29}
\eeq
Hence,  the classical trajectories  in the global coordinates are 

\beq
\tau =  \eta  \tan^{-1}  \left[  \frac{  \, \varepsilon  \sin  \theta}{ \sqrt{ \varepsilon^2 \cos^2 \theta -1}}  \right], 
\label{b30}
\eeq
whose inverse 
\beq
\theta(\tau)  = {\rm sign}( \pi - \tau) \cos^{-1} \sqrt{ \cos^2 \tau + \frac{1}{ \varepsilon^2} \sin^2 \tau}, \qquad  0 \leq \tau \leq 2 \pi, 
\label{b32}
\eeq
shows that they are periodic with  the same period $2 \pi$ for all energies.  Finally, the  position and momenta are given by

\barray 
x(\tau) & = & \sinh^{-1} ( \tan \theta(\tau))  , \qquad  0 \leq \tau \leq 2 \pi, \label{b31} \\
p(\tau)  & = &   \ell_p ( \varepsilon \cos \theta(\tau)  +  \sqrt{ \varepsilon^2 -1} \cos \tau). 
\nonumber
\earray 
Fig. 1  shows an example of a classical trajectory. The phase space contour is traversed in counterclockwise
so that the Maslov phase is $- 2 \pi$,  as for the standard harmonic oscillator. This justifies the constant $\frac{1}{2}$
in the semiclassical formula  (\ref{12})  for the energy levels.

\begin{figure}[t]
\begin{center}
\includegraphics[height= 4.0 cm]{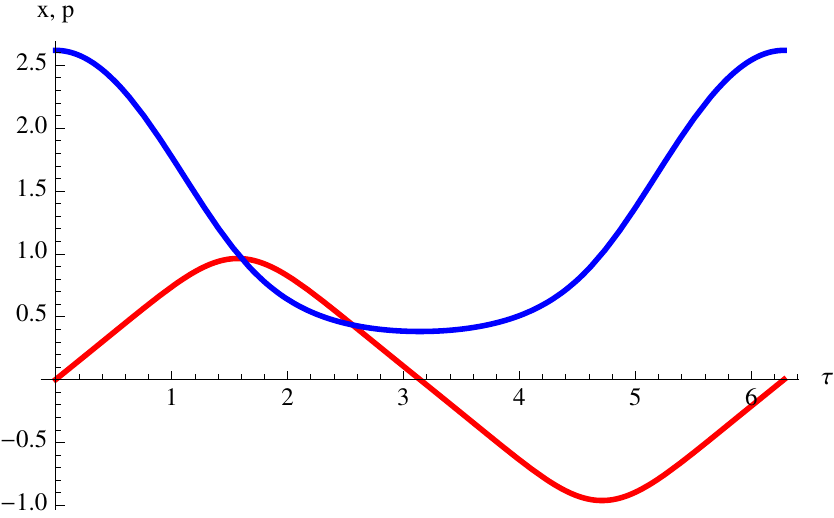}
\hspace{0.5 cm} 
\includegraphics[height= 4.5 cm]{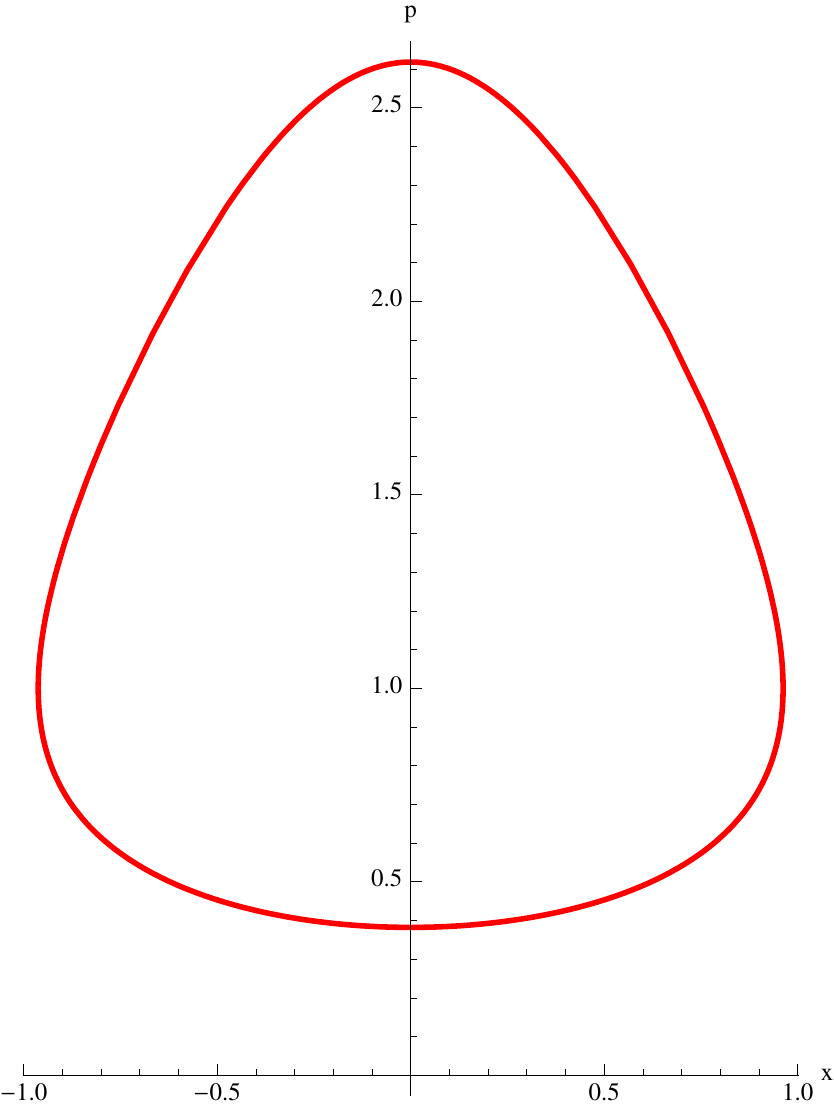}
\end{center}
Fig.1-Left: Position (red line)  and momentum (blue line)  of a classical trajectory with $E=3$ in units $w_0= \ell_p= R=1$. 
Right: Same trayectory in phase-space. 
\label{x-p}
\end{figure} 

\section{Massive Dirac fermion in AdS$_2$}
\label{sec:Dirac}

 The Dirac equation  for a massive fermion 
in 1+1 spacetime dimensions was given in eq.(\ref{e16}). 
In the case of AdS$_2$ metric (\ref{a4}), the vielbein 
and spin connection  are given by 
  
  \barray 
  e^0_0 & = &  R \, \cosh \rho, \qquad e^1_1 = R^2,  \qquad 
  \omega^{01}_0  =   - \omega^{10}_0 = \sinh \rho,  
  \label{d6}
  \earray 
  which plugged into the eq.(\ref{e16}) yield

\barray
\left( \frac{i \lambda}{\cosh \rho} + \frac{1}{2} \tanh \rho + \partial_\rho \right) \psi_+ -  i \kappa \psi_-  & = & 0,  \label{d7} \\
\left( \frac{- i \lambda}{\cosh \rho} + \frac{1}{2} \tanh \rho + \partial_\rho \right) \psi_- + i \kappa  \psi_+  & = & 0,  \nonumber 
\earray 
where we have assumed the  time dependence $e^{-  i \lambda \tau}$ for $\psi_\pm$. 
We are working in units  $\hbar = c=1$ so that  $\kappa = m R$ according to eq.(\ref{16}). 
Combining these eqs. one gets

\beq
\left( \partial_\rho^2  + \tanh \rho \, \partial_\rho + \frac{1}{4} - \kappa^2 + \frac{ i \lambda \sinh \rho}{ \cosh^2 \rho} + 
\frac{ \lambda^2 + 1/4}{ \cosh^2 \rho} \right) \psi_- = 0,
\label{d8}
\eeq
which for the function 

\beq
\phi(\rho) = (\cosh  \rho)^{ 1/2}  \psi_- (\rho)
\label{d9}
\eeq
implies 

\beq
\left( \partial_\rho^2  - \kappa^2 + \frac{ i \lambda \sinh \rho}{ \cosh^2 \rho} + 
\frac{ \lambda^2 }{ \cosh^2 \rho} \right) \phi= 0,
\label{d10}
\eeq
or  in terms of the angle $\theta$ defined in (\ref{7})

\beq
\left( \cos^2 \theta  \, \partial^2_\theta - \sin \theta \, \cos \theta \, \partial_\theta - \kappa^2 +  i \lambda 
 \sin \theta \, \cos \theta + \lambda^2 \cos^2 \theta \right) \phi = 0, 
 \label{d11}
 \eeq
and in terms of $z= e^{ 2 i \theta}$ 

\beq
\left( z ( z+1)^2 \partial_z^2 + \frac{ ( z + 1) ( 3 z +1)}{2}  \partial_z + \kappa^2 - \frac{ \lambda^2}{2}
- \frac{ \lambda + \lambda^2}{ 4 } z + \frac{ \lambda -  \lambda^2}{ 4 } z^{-1} \right) \phi = 0.
\label{d12}
\eeq
The solutions of this equation in terms of  hypergeometric functions 
regular at $z=0$, are given by

\barray
\phi^+(z)  & =  & z^{ \frac{1}{2}  -  \frac{ \lambda}{2} } (1+z)^{ -  \kappa} F( 1 - \kappa, \frac{1}{2} - \kappa - \lambda, \frac{3}{2} - \lambda, -z), 
\label{d13}  \\
\phi^-(z)  & =  & z^{  \frac{ \lambda}{2} } (1+z)^{ -  \kappa} F( \frac{1}{2}  - \kappa + \lambda, - \kappa,   \frac{1}{2} + \lambda, -z)
\nonumber 
\earray 
which using eq.(\ref{d9}) implies for the component $\psi_-$ 

\barray
\psi_-^+(z)  & =  & z^{ \frac{1}{4}  -  \frac{ \lambda}{2} } (1+z)^{  \frac{1}{2} -  \kappa} F( 1 - \kappa, \frac{1}{2} - \kappa - \lambda, \frac{3}{2} - \lambda, -z), 
\label{d13}  \\
\psi_-^-(z)  & =  & z^{  - \frac{1}{4} +  \frac{ \lambda}{2} } (1+z)^{ \frac{1}{2}  -  \kappa} F( \frac{1}{2}  - \kappa + \lambda, - \kappa,   \frac{1}{2} + \lambda, -z)
\nonumber 
\earray 
The relation between $\lambda$ and the eigenenergies $E$ of the $xp$ model can be found from the eqs.(\ref{7}), (\ref{33}) and (\ref{36})
(in units $\hbar =1$) 

\beq
e^{ - i E t} = e^{ - i \frac{E R}{2 w_0} ( \tau + \theta)} = e^{ - i \frac{\alpha}{2} ( \tau + \theta)} = e^{ - i \frac{\alpha}{2} \tau} z^{- \frac{\alpha}{4} } 
= e^{ - i \lambda  \tau} z^{- \frac{\alpha}{4} } 
\rightarrow \lambda = \frac{ \alpha}{2} 
\label{d14}
\eeq 
Using eq.(\ref{33b}) for $\alpha$ one can then easily  see that eqs.(\ref{d13}), correspond to the solutions
(\ref{50}) found for the $xp$ model.  More generally, requiring that for all values of $\kappa = m R$ (including zero) the solutions of eq. (\ref{d12}) to be single-valued leads to impose eq. (\ref{33b}) (see also \cite{Dirac}). However, recalling the mapping to the $xp$ model, we have not considered here the case $\kappa=0$ i.e $\ell_p =0$ due to in this limit the spectrum of the $xp$ model is continuous. 
Our result is an example  of the relation between the $xp$ model
and the massive Dirac equation found in section 5. 

Finally, we would like  to mention  that equation (\ref{d10}),  for the value $\lambda =2$,  coincides with the Schroedinger equation of the 
Hamiltonian of the PT-symmetric complexified Scarf II potential which is reflectionless and  possess  a hidden nonlinear supersymmetry 
\cite{CP}. It turns out that this  model is also related to the Dirac equation of a spin 1/2 particle in 
de Sitter  space \cite{L}. It would be worthwhile to explore these suggestive connections  in  further detail.


\begin{thebibliography}{99}

\bibitem{BK99} M.V. Berry,  J.P. Keating, 
``$H=xp$  and the Riemann zeros'', 
in {\em Supersymmetry and Trace Formulae: Chaos and Disorder}, 
ed. J.P. Keating, D.E. Khmelnitskii and 
I. V. Lerner, Kluwer,  1999. 

\bibitem{C99} A. Connes,  ``Trace formula in noncommutative geometry and the zeros of the Riemann zeta function'', 
 Selecta Mathematica New Series
 {\bf 5}   29,  (1999). 
math.NT/9811068.

\bibitem{ST08} G. Sierra, P.K. Townsend,  "The Landau model and the Riemann zeros", 
Phys. Rev. Lett. {\bf 101}, 110201 (2008); arXiv:0805.4079. 

\bibitem{SL11} G. Sierra,  J. Rodriguez-Laguna, 
"The $H=xp$ model revisited and the Riemann zeros".
Phys. Rev. Lett. {\bf 106}, 200201 (2011); arXiv:1102.5356


\bibitem{BK11} M. V. Berry,  J. P.  Keating, 
 "A compact hamiltonian with the same asymptotic 
mean spectral density as the Riemann zeros", J. Phys. A: Math. Theor. {\bf 44}, 285203 (2011). 

\bibitem{S11} M.  Srednicki,  "The Berry-Keating Hamiltonian and the Local Riemann Hypothesis", 
J. Phys. A: Math. Theor. {\bf 44}  305202 (2011);  arXiv:1104.1850. 

\bibitem{S11b} M.  Srednicki, 
"Nonclasssical Degrees of Freedom in the Riemann Hamiltonian", 
 Phys. Rev. Lett. {\bf 107}, 100201 (2011); arXiv:1105.2342.

\bibitem{S12} G. Sierra, "General covariant $xp$  models and the Riemann zeros", J. Phys. A: Math. Theor. {\bf 45}  055209 (2012);
arXiv:1110.3203. 

\bibitem{G12} K. S. Gupta, E. Harikumar, A.  R. de Queiroz, 
"A Dirac type xp-Model and the Riemann Zeros", arXiv:1205.6755.


\bibitem{S07a} G. Sierra, ''$H=xp$ with interaction and the Riemann zeros'',
Nucl. Phys. {\bf B 776}, 327 (2007); math-ph/0702034. 

\bibitem{TM07} J. Twamley, G. J. Milburn, ``The quantum Mellin transform'', 
New  J. Phys. {\bf 8},  328 (2006); quant-ph/0702107. 


\bibitem{J72} R. Jackiw, "Introducing Scale Symmetry" Physics Today 25 Number 1, 23 (1972).

\bibitem{AFF}  V. de Alfaro, S. Fubini and G. Furlan, "Conformal Invariance in Quantum Mechanics", Nuovo Cim. 34A, 569 (1976).


\bibitem{MS00} J. Michelson, A. Strominger, "The Geometry of (Super) Conformal Quantum Mechanics", Comm.  Math. Phys, {\bf 213}, 1 (2000).


\bibitem{BE03} S. Bellucci, A. Galajinsky, E. Ivanov and S. Krivonos, "AdS$_2$/CFT$_1$, canonical transformations and superconformal mechanics" Phys.  Lett.  B, {\bf 555}, 99 (2003).


\bibitem{S98} A. Strominger, "AdS$_2$ Quantum Gravity and String Theory" JHEP 9901:007 (1999); hep-th/9809027.


\bibitem{Ch11} C. Chamon, R. Jackiw, S.-Y. Pi, L. Santos, "Conformal quantum mechanics as the CFT$_1$ dual to AdS$_2$",
Phys.  Lett.  B {\bf 701}, 503 (2011), arXiv:1106.0726. 

\bibitem{JP} R. Jackiw, S.-Y. Pi,  "Conformal Blocks for the 4-Point Function in Conformal Quantum Mechanics", 
Phys. Rev. D {\bf 86}, 045017 (2012);  arXiv:1205.0443. 


\bibitem{GA08} A. Galajinsky, "Particle dynamics on AdS$_2\, \times$ S$_2$ background with two-form flux", Phys. Rev. {\bf 78}, 044014 (2008). 

\bibitem{PL03} C. Leiva, M. S. Plyushchay.
"Conformal symmetry of relativistic and nonrelativistic systems and AdS/CFT correspondence", Annals Phys. {\bf 307} 372-391 (2003). 


\bibitem{PL09} F. Correa, V. Jakubsky and M. S. Plyushchay, "Aharonov-Bohm effect on AdS$_2$ and nonlinear supersymmetry of reflectionless Poschl-Teller system", Annals Phys. {\bf 324}  1078-1094 (2009). 


\bibitem{AS} M. Abramowitz and I. A. Stegun,  {\em Handbook of mathematical functions}, 
Dover, New York 1974.  (1972).  


\bibitem{ads} O. Aharony, S.S. Gubser, J. Maldacena, H. Ooguri, Y. Oz, 
"Large N Field Theories, String Theory and Gravity", Phys. Rep. {\bf 323} 183 (2000); 
arXiv:hep-th/9905111.


\bibitem{strominger2} A. Strominger, "A matrix model for AdS$_2$", JHEP {\bf 03} (2004) 066; arxiv:hep-th/0312194. 

\bibitem{Dirac} O. Aharony and A. Patir, "The Conformal Limit of the 0A MatrixModel and String Theory on AdS$_2$",
JHEP 0511:052 (2005); arXiv:hep-th/0509221. 


\bibitem{Villa} V.  M. Villalba, 
"Dirac equation in some homogeneous space-times, separation of
variables and exact solutions",   Mod. Phys. Lett. A {\bf 8} 
2351 (1993);   arXiv:gr-qc/9306019.

\bibitem{BG} B. Basu-Mallick, K.S. Gupta, Phys. Lett. {\bf A}  292, 36 (2001);  hep-th/0109022.

\bibitem{susy} F. Correa, M. A. del Olmo, M. S. Plyushchay, 
"On hidden broken nonlinear superconformal symmetry of conformal mechanics and nature of double nonlinear superconformal symmetry",
Phys. Lett. {\bf B}  628, 157  (2005); arXiv:hep-th/0508223. 


\bibitem{CP} F. Correa, M.  S. Plyushchay, 
"Self-isospectral tri-supersymmetry in PT-symmetric quantum systems with pure imaginary periodicity", 
Annals Phys. 327,  1761 (2012); arXiv:1201.2750. 

\bibitem{L}  P. Lagogiannis, A. Maloney, Y. Wang, "Odd-dimensional de Sitter space is transparent";  arXiv:hep-th/1106.2846. 




\end{thebibliography}
\end{document}